\documentclass[a4paper,aps,prb,twocolumn,floatfix,showpacs]{revtex4}%
\usepackage{graphicx}
\usepackage{epsfig}
\usepackage{amsmath}
\usepackage{amssymb}
\usepackage{amsfonts}%
\begin{document}
\title{Electronic thermal conductivity of disordered metals}
\author{Roberto Raimondi}
\affiliation{Dipartimento di Fisica "E. Amaldi," Universit\`a di Roma Tre, Via della Vasca Navale 84, 00146 Roma, Italy}
\author{Giorgio Savona}
\affiliation{Dipartimento di Fisica "E. Amaldi," Universit\`a di Roma Tre, Via della Vasca Navale 84, 00146 Roma, Italy}
\author{Peter Schwab}
\affiliation{Institut f\"ur Physik, Universit\"at Augsburg, 86135 Augsburg Germany}
\author{Thomas L\"uck}
\affiliation{Institut f\"ur Physik, Universit\"at Augsburg, 86135 Augsburg Germany}

\begin{abstract}
We calculate the thermal conductivity of interacting electrons in disordered metals.
In our analysis we point out that the interaction affects thermal transport through two distinct mechanims,
associated with quantum interference corrections and energy exchange of the quasi particles with 
the electromagnetic environment, respectively. The latter is seen to
lead to a violation of the Wiedemann-Franz law. 
Our theory predicts a strong enhancement of the Lorenz ratio $\kappa /\sigma T $ 
over the value which is predicted by the
Wiedemann-Franz law, when the electrons
encounter a large environmental impedance. 
\end{abstract}
\date{\today}
\maketitle


The Wiedemann-Franz law relates the electronic thermal conductivity, $\kappa$,
and the electrical conductivity, $\sigma$, 
and states that the Lorenz ratio  $L=\kappa /\sigma T$ is a universal constant given by
$L= \pi^2 k_B^2/ 3 e^2$.
In this equation  $k_{B}$ is the Boltzmann constant, $-e$ the electron charge, and $T$
the temperature. 
The validity of the Wiedemann-Franz law relies mainly on a
single-particle description of the transport properties,   on the Fermi
statistics of the charge carriers and on the assumption of purely elastic scattering\cite{chester1961}.
In a Fermi-liquid, one expects that this law
still holds at low enough temperature,
when the quasi-particles cannot exchange energy during collisions
\cite{langer1962}.
Deviations from the Wiedemann-Franz law as recently observed in the normal state of
a copper-oxide superconductor have thus been interpreted as an evidence for the breakdown
of Fermi-liquid theory 
\cite{hill01}.

The effects of Coulomb interaction on the electrical transport at low temperature
can be, broadly, grouped
in two main types. From one side, transport implies adding charges to
a conductor. This has an energy cost that depends on the size and shape of the
conductor itself. 
For example in tunnel junctions the energy transfer
between quasi-particles and the electrodynamical environment
causes the Coulomb blockade phenomena.
On the other side Coulomb
interaction leads in a disordered conductor to an additional source of random
scattering that interferes with the scattering from the impurities. This is a quantum
effect and depends on the details of charge diffusion.

In the 1980s 
these quantum interferences
were
shown to lead to corrections to the electrical conductivity
beyond the standard Fermi-liquid results. 
It turned out that these corrections may, in fact,
be incorporated into a scale-dependent renormalization of the Landau Fermi-liquid 
parameters \cite{altshuler1979,altshuler1983,finkelstein1983,
castellani1984,castellani1986,castellani1986b}. 
Whithin this framework, Castellani and co-workers \cite{castellani1987}
demonstrated that the scale-dependent corrections to the thermal conductivity
are the same as the corrections to the electrical conductivity. This led them to
conclude that the Wiedemann-Franz law is valid up to the metal-insulator transition.

This conclusion was challenged by Livanov et al.~\cite{livanov1991} who
found for a two-dimensional system with long range Coulomb interaction
additional contributions to the thermal conductivity.
Recently, the issue has been reexamined by Niven and Smith \cite{niven2003},
who also concluded that the Wiedemann-Franz law is violated.

In this paper we 
study the problem by means of the quasiclassical Green's function approach,
which 
has proved to be a 
powerful tool in describing 
the dynamical properties of superconductors \cite{langenberg1986} and the
transport in hybrid mesoscopic structures \cite{lambert1998}.
Recently it was further  demonstrated that both Coulomb blockade phenomena 
and quantum interference corrections to the charge transport can be conveniently 
described within this theoretical framework \cite{schwab2003}.
Advantages of the method are that it is not restricted to the linear response
regime, and often provides 
more compact derivations than the standard diagrammatic techniques.

A perturbative calculation of the thermal conductivity, besides confirming Ref.~\cite{niven2003},
allows us to clarify the origin of the apparent discrepancies in the literature.
To do so we separate the different physical mechanisms by
their different range of exchanged energies and relevant length scales.  
For instance, the quantum interference effects occur over
distances from  the mean free path up to the thermal diffusion length $\sqrt{\hbar D / k_B T}$, 
and imply energy exchanges larger than
the temperature $T$.
These yield corrections which are  logarithmically divergent 
and can be readily related
to the scale-dependent renormalization of the electrical conductivity. 
Here the temperature acts as an infrared cutoff.
The interaction effects responsible
for the deviations from the Wiedemann-Franz law are 
associated with the long-range part of the Coulomb interaction
and their singular behavior has the temperature as the upper cutoff. 
We successively concentrate on the long-range part of the Coulomb interaction. 
In particular 
we predict a sizable enhancement of the Lorenz ratio when the sheet 
resistance is of the order $h/e^2$ in the case of a two-dimensional electron system,
and a strong enhancement of the Lorenz ratio for thin metallic wires when
the total resistance of the wire is larger than $h/e^2$.

From now on we use units where $\hbar =k_B=1$, but we put back the constants in the final results.
We start with a brief introduction of the 
quasiclassical formalism. For a more detailed description 
we defer the reader to Ref.~\cite{rammer1986}. 
Whithin this formalism the  short-distance behavior of the electron Green's function is
taken into account in an averaged way from the outset by introducing 
the quasiclassical Green's function, which solves the Eilenberger equation~\cite{eilenberger1968}
\begin{equation}
\left[ {\partial}_{t_1}+ \partial_{t_2} +v_{F}{\hat{\mathbf{p}}\cdot \partial}_{\mathbf{x}}\right]
\check{g}_{t_1 t_2}\left( \mathbf{x},{\hat{\mathbf{p}}}\right)  =-{\rm i}
\left[
\check{\Sigma}\left(  \mathbf{x},{\hat{\mathbf{p}}}\right)
,\check{g}\left(  \mathbf{x},{\hat{\mathbf{p}}}\right)  \right] .
\label{1,1}
\end{equation}
In contrast to the  Dyson equation for the ordinary Green's function, 
the Eilenberger equation for $\check{g}$
is homogeneous and requires a normalization condition, which can be chosen of the
form  ${\check{g}\check{g}}={\check{1} }$.
The Green's function has a two-by-two matrix structure in Keldysh space,
\begin{equation}
\label{matrixstructure}
\check g = 
\left(
\begin{array}
[c]{cc}
g^R  & g^K   \\
0  & g^A 
\end{array}
\right)
.\end{equation}
Matrix products imply both summation and integration over Keldysh indices and time variables,
respectively. We recall that, whereas the  diagonal components of $\check{g}$
describe the spectral properties of the system,
the off-diagonal Keldysh component  carries information
about the distribution function.  In this respect, the Keldysh component of Eq.~(\ref{1,1})
is the quantum analog of the Boltzmann equation.

Impurity scattering is introduced by means of the standard
white-noise random potential and is described by the 
self-energy in the self-consistent Born-approximation as
\begin{equation}
\label{born}
\check{\Sigma}_{t_1 t_2}({\bf x})=-\frac{{\rm i} }{2\tau}\int\frac{d{\hat{\mathbf{p}}}}{\Omega_d}
\check{g}_{t_1 t_2}
\left(  \mathbf{x},{\hat{\mathbf{p}}}\right) 
,\end{equation}
where $\tau$ is the elastic scattering time and $\Omega_d$ is the $d$-dimensional solid angle.

The charge and heat current densities have the form
\begin{equation}
\left( \begin{array}{c}
 {\bf j}_e({\bf x}, t)  \\  
 {\bf j}_Q({\bf x}, t)
 \end{array}
\right)
 = 
- \frac{N_{0}}{2}v_F {\int}
\frac{d \hat {\bf p} }{\Omega_d }
{\hat{\mathbf{p}}}
{\int}d\epsilon 
\left(  \begin{array}{c }- e  \\ \epsilon \end{array}  \right)
 g^{K}\left(  \mathbf{x} ,t;{\hat{\mathbf{p}}},\epsilon\right)  
 \label{6}
,\end{equation} 
where $N_0$ is the density of states at the Fermi energy, $t=(t_1+t_2)/2$ and $\epsilon$ 
corresponds to the Fourier transform of the relative time $t_1 -t_2$.

In the dirty limit, the variation of the Green's function  is on
 space and time scales larger than the elastic mean free path $l=v_F \tau$
and scattering time $\tau$, respectively. 
In this limit one
may expand the Green's function $\check{g}$ in spherical harmonics and keep
 the $s$- and $p$-wave components only,
$ \check{g}( \hat {\bf p})=\check{g}_{s}+\hat{\mathbf{p}}\check{g}_{p}+... $
The Eilenberger equation is then replaced by 
( $D=v_F^2 \tau /d$) the Usadel equation\cite{usadel1970},
\begin{equation}
\label{usadel}
\partial_t \check g_s({\bf x}) - D \partial_{\bf x}
\left[  {\check g}_s\partial_{\bf x} \check g_s  \right] = 0
\end{equation}
which is the analogous of  the saddle-point
condition in the non-linear $\sigma$-model matrix field theory\cite{finkelstein1983}.
As a result,
the currents are expressed in terms of the
$s$-wave component of the Green's function:
\begin{eqnarray} \label{eq6}
\mathbf{j}_e(\mathbf{x},t) &= & 
-\frac{eN_{0}D}{2} \int d\epsilon  
\left(  \check g_{s}
\partial_{\mathbf{x}}\check g_{s}\right)^K ,
\\ \label{eq7}
\mathbf{j}_Q(\mathbf{x},t) &= & 
\frac{N_{0}D}{2} \int d\epsilon \, \epsilon  \, 
\left(  \check g_{s}  \partial_{\mathbf{x}}\check g_{s} \right)^K.   
\end{eqnarray}

As a simple application of the formalism we derive the Drude formula for the
electrical and thermal conductivity. 
In the absence of interactions the Green's function $\check{g}$
reads:
\begin{equation}
{\check{g}}_{s}(\mathbf{x},t; \epsilon)=
\left(
\begin{array}
[c]{cc}
1  & 2F \\
0  & -1
\end{array}
\right)\label{8}
.\end{equation}
Near local equilibrium with a local temperature $T({\bf x})$ and
chemical potential $\mu({\bf x})$ the function $F$ is given by
\begin{equation}
F =\tanh\left(
\frac{\epsilon - \mu({\bf x}) }{2T\left(  \mathbf{x}\right)  }\right) \label{disfunc}
,\end{equation}
from which the Drude expressions for both electrical and heat currents are found 
\begin{eqnarray}
{\bf j}_e &=& 2 e^2 D N_0  ( \nabla \mu /e ) \\
{\bf j}_Q &=& \frac{\pi^2}{3}k_B^2 2 N_0 D  T (- \nabla T )  
,\end{eqnarray}
and in particular the Wiedemann-Franz law holds.

To include the effects of Coulomb interaction, 
we introduce\cite{schwab2003} a Hubbard-Stratonovich matrix field
\begin{equation}
\check{\phi}=\left(
\begin{array}
[c]{cc}
\phi_{1}  & \phi_{2}\\
\phi_{2}  & \phi_{1}
\end{array}
\right)  \label{9}
\end{equation}
whose fluctuations describe the retarded, advanced, and Keldysh
components of the screened Coulomb interaction
\begin{equation}
-{\rm i} e^{2}\left(
\begin{array}
[c]{cc}
\left\langle \phi_{1}\phi_{1}\right\rangle  &  \left\langle \phi_{1}\phi
_{2}\right\rangle \\
\left\langle \phi_{2}\phi_{1}\right\rangle  & \left\langle \phi_{2}\phi
_{2}\right\rangle
\end{array}
\right)  =\frac{1}{2}\left(
\begin{array}
[c]{cc}
V^{K} &  V^{R}\\
V^{A} &   0
\end{array}
\right)  .\label{10}
\end{equation}
In the presence of the field $\check \phi$, one first adds  a term  
${\rm i} e [\check \phi, \check g] $
 to the right-hand side of the
Eilenberger (\ref{1,1}) or the Usadel equation (\ref{usadel}).
Secondly, the resulting solution $\check{g}\left[  \phi\right]$ is averaged
over the fluctuations of $\check{\phi}$ according to Eq.~(\ref{10}).
In analogy to the non-interacting case (Cf.\ Eq.~(\ref{8})),
one can define the distribution function
in the presence of interactions via the relation between the Keldysh and the retarded, advanced
components of the Green's function $\langle g^K \rangle = \langle g^R \rangle F - F \langle g^A \rangle$.
We further assume a system which is -- with the exception of a weak temperature gradient  --
translational invariant.
Then it is convenient to expand the distribution function as
\begin{equation}
F_{\epsilon-\omega}({\bf x}_1) \approx F_{\epsilon-\omega}({\bf x}) - \partial_T F_{\epsilon-\omega}({\bf x})
\nabla T \cdot ({\bf x}_1 - {\bf x})
\end{equation}
and to Fourier transform from real to momentum space.
The correction to the thermal current is finally obtained as
$\delta {\bf j}_Q = \delta {\bf j}_Q^a + \delta {\bf j}_Q^b$ with
\begin{eqnarray}
\delta {\bf j}_Q^a  & =  & {DN_0}\nabla T \int d \epsilon \, \epsilon \,
\int \frac{d \omega }{2 \pi}
 \partial_T(F_{\epsilon-\omega}({\bf x}) F_\epsilon({\bf x}) ) \nonumber\\
\label{eq20}
&&\times \, {\rm Im} \, \sum_q \frac{1}{( -{\rm i}\omega + Dq^2)^2 } V^R_\omega ({\bf q}) 
\end{eqnarray}
and
\begin{eqnarray}
\label{eq21}
\delta {\bf j}_Q^b & = & {D N_0} \nabla T \int d \epsilon \, \epsilon \,
\int \frac{d \omega}{2 \pi}
 F_\epsilon({\bf x})\partial_T F_{\epsilon-\omega}({\bf x}) \nonumber\\
&& \times \frac{4}{d} \, {\rm Im} \,  \sum_q \frac{Dq^2}{( -{\rm i }\omega + Dq^2)^3 } V^R_\omega({\bf q})
,\end{eqnarray}
where $d$ is the dimension of the system under consideration.
Our result, Eqs.~(\ref{eq20}-\ref{eq21}), for the thermal current is 
equivalent to the thermal conductivity
found in \cite{niven2003} by using the diagrammatic method and the Matsubara technique.
We notice that the diffusive pole appearing in Eqs.~(\ref{eq20}-\ref{eq21})
originates from the Usadel equation (\ref{usadel}).

Using the relation $F_\epsilon F_{\epsilon-\omega}=1- (  F_\epsilon-F_{\epsilon-\omega} ) B (\omega /2T)$
with $B (x)=\coth(x)$
allows to evaluate the $\epsilon$-integrations in Eqs.~(\ref{eq20}-\ref{eq21})
 with the result
\begin{align}
\label{eq23}
 & \int d \epsilon\, \epsilon \, \partial_T(F_\epsilon F_{\epsilon - \omega})  =  
-\omega^2 \partial_T B\left (\frac{\omega}{2 T}\right),  \\
 & \int d \epsilon\, \epsilon \, F_\epsilon \partial_T F_{\epsilon - \omega }  = 
-\frac{2 \pi^2 T}{3} \partial_\omega [\omega B\left (\frac{\omega}{2 T}\right)]
+ \frac{\omega^3}{3 T }
 \partial_\omega  B\left (\frac{\omega}{2 T}\right)\label{eq24}
.\end{align}
From Eqs.~(\ref{eq23}) and (\ref{eq24}) we observe that 
$\delta {\bf j}^a_Q $ is dominated by diffusive modes of frequency $ |\omega |  < T$, whereas 
modes with frequencies $| \omega | > T$ give the dominant contribution to $\delta {\bf j}^b_Q$.
To appreciate the role played by the
different  frequency ranges  we begin by evaluating the current in two dimensions.
The retarded component of the dynamically screened  Coulomb interaction
reads
\begin{equation} 
V^{R}\left(  \mathbf{q}, \omega\right) \approx \frac{1}{2N_0}\frac{\kappa_{2d}}{q} \frac{ -{\rm i} \omega + D q^2}
{- {\rm i} \omega+ D\kappa_{2d} q }, \label{10,1}
\end{equation}
where 
$\kappa_{2d} = 4 \pi e^2 N_0 $ is the screening vector in two dimensions.
By considering  first  $\delta {\bf j}_Q^b$
one notices that the momentum integration to be performed 
is identical to the momentum integral in the correction to the electrical conductivity 
\cite{altshuler1979,altshuler1983,finkelstein1983,castellani1984}, i.e.,
the integration is logarithmically  divergent in the ultraviolet and must be
cutoff with the diffusive condition $Dq^2\tau <1$.
In the $\omega$-integration there is a minor difference at low frequencies $ |\omega | < T$ due to
the second term on the right-hand side of Eq.~(\ref{eq24}). 
In two dimensions, with logarithmic accuracy,
this difference is negligible and one has
\begin{eqnarray}
\delta {\bf j}^b_Q &\approx  & \frac{\pi^2}{3} \frac{T}{e^2} \delta \sigma (- \nabla T)
,\end{eqnarray}
where $\delta \sigma = -e^2/(2\pi^2) \ln(1/ T\tau)$ is the interaction correction to 
the electrical conductivity and $T\tau < 1$.
The other contribution to the thermal current, $\delta {\bf j}_Q^a$,
does not depend on the ultraviolet cutoff $1/\tau$, 
\begin{eqnarray}
\delta {\bf j}^a_Q & \approx & 
-  D\nabla T  \int_0^T\frac{d \omega}{ 2\pi } \omega
   \int \frac{d^2 q}{(2 \pi )^2}
 \nonumber \\
&& \times {\rm Im} \left( \frac{1}{-{\rm i}\omega + Dq^2}
\frac{\kappa_{2d}}{q}
\frac{1}{-{\rm i}\omega + D q \kappa_{2d}} \right) 
\end{eqnarray}
since the temperature acts as an upper cutoff in the frequency integration.
In contrast, in the limit of good metallic screening
when $\kappa_{2d} \to \infty$, the integration becomes infared divergent.
By combining the two contributions we
finally write the expression for the thermal conductivity in a form which shows that,
although  the Wiedemann-Franz law is violated,
\begin{equation}
\kappa = \frac{\pi^2}{3}\frac{k_B^2 T}{e^2} \left( \sigma + \delta  \sigma + 
\frac{1}{2} \frac{e^2}{\pi h} \ln(\hbar D \kappa_{2d}^2 / k_B T )
\right)  \label{29}
,\end{equation}
the integration of diffusive modes in the region $T< Dq^2, \omega <\tau^{-1}$ yields the same 
scaling equations for $\sigma $ and $\kappa$,
\begin{equation}
\frac{d \ln \sigma}{d \ln l } = \frac{d \ln \kappa}{d \ln l }
,
\end{equation}
so that the apparent discrepancies in the literature are no contradiction.

We observe that in the last term of Eq.~(\ref{29}), 
responsible for the violation of the Wiedemann-Franz law,
only the extreme long wavelength modes of the dynamically screened Coulomb interaction
with  $Dq^2< |\omega |< T$ are relevant, cf.\ Eqs.~(\ref{eq20}) and
(\ref{eq23}). It has been shown in Ref.\cite{kopietz1998}  that these
can be summed to all orders and in the end modify the Green's function like a gauge factor,
\begin{equation}
\check{g}_{t_{1}t_{2}}\left( {\bf x}; \hat{{\bf p} } \right)  =
e^{{\rm i} \check{\varphi}\left( {\bf x}, t_{1}\right)  }
\check{g}_{_{0}t_{1}t_{2}}({\bf x};\hat{\bf p})
e^{-{\rm i}\check{\varphi}\left( {\bf x}, t_{2}\right)  },\label{35}
\end{equation}
where $ \partial_t \check \varphi({\bf x}, t) = e \check \phi({\bf x}, t)$.
Whereas the gauge factors drop in the expression for the electrical
current, i.e.\ the long wavelength modes of the Coulomb interaction do not modify the electrical
conductivity, they survive
in the heat current to yield
\begin{align}
 \mathbf{j}_{Q}&= - \frac{ N_{0} }{2}
\int \frac{d \hat {\bf p} }{\Omega_d} v_F \hat {\bf p}
\int d \epsilon \, \Big\{ \epsilon  \,  g^K_{0}({\bf x}, t; \hat {\bf p}, \epsilon) \nonumber \\
&- \frac{1}{2} e
 \left( \langle \check \phi(t) \check g({\bf x},t;\hat{\bf p},\epsilon ) + 
 \check g({\bf x},t;\hat{\bf p},\epsilon) \check \phi(t) \rangle  \right)^K \Big\} \label{eq30}
.\end{align}
Eq.~(\ref{eq30}) makes clear the physical origin of the violation of the
Wiedemann-Franz law. While the first term on the right-hand side reproduces 
the non-interacting contribution to the thermal current, the second
may be interpreted as the effect of the time dependent fluctuations of the quasi-particle energy 
in the presence of an electromagnetic environment.
Indeed, the extra heat current is proportional to the correlation of voltage 
and current fluctuations in the system,
$ \delta {\bf j}_Q = \langle \phi_1({\bf x},t) \delta {\bf j}_e({\bf x},t ) \rangle$,  
which then leads to the strikingly simple result
\begin{equation}
{\bf j}_Q = \frac{\pi^2}{3} \frac{k_B^2 T }{e^2} \sigma (- \nabla T ) -  \frac{1}{2} \sigma \nabla 
\langle \phi_1({\bf x}, t) \phi_1({\bf x}, t) \rangle
.\end{equation}
Notice that due to the linear current-voltage characteristics
of the system under consideration
only the first order in the Coulomb interaction contributes to the
heat current.
By using the fluctuation dissipation theorem (or equivalently Eq.~(\ref{10}))
\begin{eqnarray}
\langle \phi_1({\bf x}, t) \phi_1({\bf x}, t) \rangle &=&  
-\frac{1}{e^2}  \int \frac{d \omega }{2 \pi} 
B\left(\frac{\omega}{ 2T({\bf x})}\right) 
 \sum_q {\rm Im} V^R({\bf q}, \omega )  \nonumber \\
&=& \int \frac{d \omega }{2 \pi} 
B\left(\frac{\omega}{ 2T({\bf x})}\right)\, \omega \,
 {\rm Re }Z(\omega) 
\end{eqnarray}
direct contact can be made with the conventional perturbation theory, i.e.\ with $\delta {\bf j}_Q^a$ in Eq.~(\ref{eq20}).

Instead of parameterizing the local voltage fluctuations in terms of an interaction $V^R({\bf q}, \omega )$ 
we will in the following parameterize them
in terms of the impedance of the local electromagnetic environment, $Z(\omega)$. 
By doing so, the thermal conductivity reads
\begin{equation}
\kappa = \frac{\pi^2}{3} \frac{k_B^2 T}{e^2} \sigma  
 + \frac{ \sigma  k_B }{e^2}
\int d E \left( \frac{E/ 2 k_B T}{\sinh( E/2 k_B T ) } \right)^2   
\frac{{\rm Re }Z(E/\hbar)}{h/e^2 }
,\end{equation}
where for clarity we put back the $\hbar$ and $k_B$.
We will now discuss three different examples for the impedance $Z$.
The  simplest situation consists of a purely ohmic environment where $Z(E/\hbar) = R$.
The thermal conductivity is found linear in the temperature, strong deviations from the Wiedemann
Franz law are found when the environmental resistance is of the order 
of the resistance quantum $h/e^2$ or larger.
The explicit result  is
\begin{equation} \label{eq36}
\kappa = \frac{\pi^2}{3} \frac{k_B^2T}{e^2} \sigma ( 1 + {2R}/{h/e^2} )
.\end{equation}
From the retarded Coulomb interaction as given in Eq.~(\ref{10,1}) we determine the impedance of a thin film as 
${\rm Re} Z(E/\hbar) = (1/{4 \pi \sigma}) \ln( \hbar D \kappa_{2d}^2 k_B T/ E^2 )$.
Due to the weak logarithmic energy dependence of the impedance the thermal conductivity is to good accuracy 
obtained from Eq.~(\ref{eq36}) with $R= {\rm Re} Z (k_B T/\hbar)$.
As a third example we consider a $RC$-transmission line, as a model of a gated wire.
The impedance is $ {\rm Re} Z  = \frac{1}{2}\sqrt{R_0/2 |\omega| C_0 }$, where $R_0$ and $C_0$ are
the resistance and capacitance per unit length. 
We find a contribution to the thermal conductivity which is proportional 
to the square root of the temperature,
\begin{equation}
\label{eq35}
\kappa =\frac{\pi^2}{3} \frac{k_B^2T }{e^2} \sigma  + 2.456
 \frac{k_B}{e^2} \sigma   
\frac{ \sqrt{\hbar k_B T R_0/ C_0 } }{h/e^2} 
,\end{equation}
with $2.456$ the approximate numerical value for $3\zeta({3}/{2})
\Gamma({3}/{2})/2^{3/2}$. 

In summary we calculated the thermal conductivity of disordered metals.
In the two-dimensional electron system
the scaling equations for the thermal and the electrical conductivity are the same, 
nevertheless the Wiedemann-Franz law does not hold. 
The deviations from the Wiedemann-Franz law are comparable in size to the localization effects.
It is interesting to note that this is in qualitative agreement with observations made in the cuprates \cite{hill01,proust2002}:
The resistivity of PCCO in Ref.\cite{hill01} shows a well pronounced low temperature anomaly which
has been attributed to localization effects, and at the same time the low temperature heat conductivity is larger
than what would be expected from the Wiedemann-Franz law. 
In the low temperature resistivity of Tl-2201 in Ref.\cite{proust2002} 
no indications of localization effects are seen and the Wiedemann-Franz law is perfectly obeyed within the experimental accuracy.
Quantitatively on the other hand the agreement of our theory with Ref.\cite{hill01} remains poor, 
since the sheet resistance was estimated as $R_\square \approx h/(60e^2)$ from which 
we expect a much smaller enhancement of the heat conductivity than observed experimentally.
By measuring the Lorenz ratio in a gated film or wire as function of the
gate capacitance, it should be possible to test our predictions experimentally.

R.R.\ and G.S.\ acknowledge partial financial support from MIUR under grant
COFIN2002022534.
We acknowledge valuable discussions with U.\ Eckern, C.\ Castellani, and C.\
Di Castro.

\end{document}